\documentclass[twocolumn,showpacs,preprintnumbers,amsmath,amssymb]{revtex4}

\usepackage{graphicx}
\usepackage{dcolumn}
\usepackage{bm}

\begin{document}


\title{Direct imaging of the structural change generated by dielectric breakdown in MgO based magnetic tunnel junctions}

\author{A. Thomas}
 \email{athomas@physik.uni-bielefeld.de}
 \homepage{http://www.spinelectronics.de}
\author{V. Drewello}%
\author{M. Sch\"afers}%
\author{A. Weddemann}
\author{G. Reiss}
\affiliation{Bielefeld University, Department of Physics, Thin films and physics of nanostructures, 33501 Bielefeld, Germany}

\author{G. Eilers}
\author{M. M\"unzenberg}
\author{K. Thiel}
\author{M. Seibt}

\affiliation{IV. Physikalisches Institut der Georg-August-Universit\"at G\"ottingen and Sonderforschungsbereich 602, Friedrich-Hund-Platz 1, D-37077 G\"ottingen, Germany}



\date{\today}

\begin{abstract}
MgO based magnetic tunnel junctions are prepared to investigate the dielectric breakdown of the tunnel barrier. The breakdown is directly visualized by transmission electron microscopy measurements. The broken tunnel junctions are prepared for the microscopy measurements by focussed ion beam out of the junctions characterized by transport investigations. Consequently, a direct comparison of transport behavior and structure of the intact and broken junctions is obtained. Compared to earlier findings in Alumina based junctions, the MgO barrier shows much more microscopic pinholes after breakdown. This can be explained within a simple model assuming a relationship between the current density at the breakdown and the rate of pinhole formation.
\end{abstract}

%
\maketitle
In 1975 the tunnel magnetoresistance effect (TMR) in ferromagnet/ insulator/ ferromagnet systems was discovered by Julli\`ere \cite{PL1975V54AS225}. In 1995 significant TMR was found at room temperature \cite{PRL1995V74S3273,MMM1995V139S231}. Since then, magnetic tunnel junctions (MTJ) became potential candidates for magnetic random access memory (MRAM) \cite{sci1998V282S1660} and have already replaced the giant magnetoresistance \cite{PRB1989V39S4828,PRL1988V61S2472} read heads in hard discs.

The dielectric breakdown of the junctions, which is of major importance for their reliability, has been investigated for the formerly used Alumina based  \cite{JAP2001V89S8038, JAP2002V91S4348, JAP2001V89S586, apl1998V73S2363} as well as the recently introduced MgO based MTJs \cite{Khan2008} due to their possible use in electronic devices. These studies were generally done by analyzing the transport properties of the MTJs \cite{JAP2001V89S8038, JAP2002V91S4348, JAP2001V89S586, Khan2008} and/ or indirect imaging of the proposed breakdown mechanism \cite{apl1998V73S2363}, generally single pinholes in the insulating barrier. Direct investigations of intact MTJs were also performed \cite{apl2006V89S042506,jap2007V101S013907}.

Here, we try to directly image the broken barriers of the tunnel junctions. In order to do this, we prepared MTJs and performed standard transport measurements. Then, one half of the junctions was stressed by high voltages to induce the dielectric breakdown. Finally, slices were cut out of the broken and intact junctions by focussed ion beam (FIB) and investigated with transmission electron microscopy (TEM). This allows to directly compare structure and transport properties of the intact and broken samples and to compare the findings with those of alumina based MTJs. A simple model is presented to explain the behavior. 
%
The magnetic tunnel junctions are prepared in a magnetron sputter system with a base pressure of $1\times10^{-7}$\,mbar. The layer stack is Ta 5/ Cu 30/ Ta 5/ Cu 5/ Mn$_{83}$Ir$_{17}$ 12/ Co$_{40}$Fe$_{40}$B$_{20}$ 4/ Mg 0.5/ MgO 1.5/ Co$_{40}$Fe$_{40}$B$_{20}$ 6/ Ta 5/ Cu 40/ Au 30 (all values in nm) on top of a thermally oxidized (50\,nm) silicon (100) wafer. To activate the exchange biasing and for the crystallization of the MgO barrier, the layer stack is annealed after sputtering at 623\,K for 60 minutes in a magnetic field of 6500\,Oe. The stack is patterned by optical lithography and ion beam etching. The junction sizes are between $\rm 7 \times 7$ and $\rm 22.5 \times 22.5\,\mu m^2$. All measurements are done by conventional two probe technique.

The transmission electron microscopy samples were prepared by FIB with a FEI NOVA NANOLAB 600, which allows sample preparation out of any desired region of a structure and permits a target preparation of a TMR device. A 0.5\,$\mu m$ thick Pt strap was ion-beam deposited on top of the area of interest in order to protect the specimen from Ga$^+$ implantation and damage during FIB milling. A 1.5\,$\mu m$ thick cross-section was cut out of the TMR element and milled with a Ga$^+$ beam at 30\,kV. Finally, the last milling step was done at 5\,kV and an angle of incidence set to 7$^\circ$, in order to reduce the thickness of amorphous surface layers.

The TEM work was done using a Philips CM200-FEG-UT operated at an acceleration voltage of 200\,kV. The microscope has a point resolution of 0.19\,nm and an information limit of 0.11\,nm. Images were recorded using a GATAN SSC charge-coupled device (CCD) (Model 694).

%
%
\begin{figure}
\includegraphics[width=80mm]{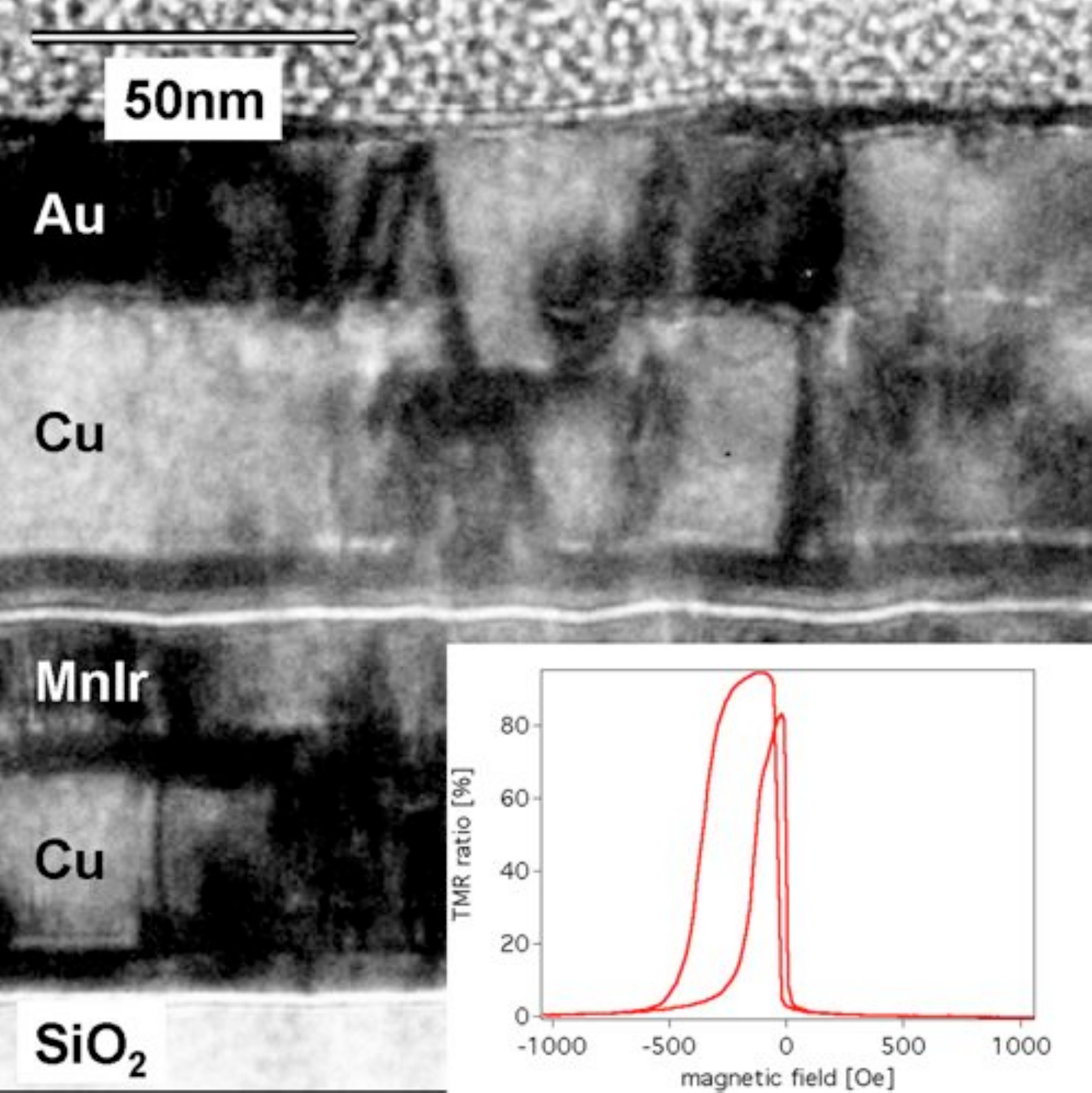}
\caption{\label{fig:intact_overview} Low-magnification TEM micrograph in cross-section geometry obtained from a working TMR element grown on thermally oxidized silicon (SiO$_2$) with some of the deposited layers indicated. On this scale, the MgO tunneling barrier looks homogeneous. The inset shows the corresponding $R$ vs.\ $H$ loop of the same junction before focussed ion beam thinning.}
\end{figure}
Figure \ref{fig:intact_overview} shows a TEM image of an MTJ. The small inset shows the magnetoresistance loop of the same junction before it was sliced by FIB, a magnetoresistance ratio of 95\,\% was determined. The TEM micrograph shows the layer stack on the thermally oxidized silicon substrate ($SiO_2$). The layer sequence mentioned above is clearly revealed. It should be noted that on this scale the MgO tunneling barrier appears to be homogeneous in thickness due to the correlated roughness of the two interfaces to the CoFeB top and bottom electrodes. 

\begin{figure}
\includegraphics[width=80mm]{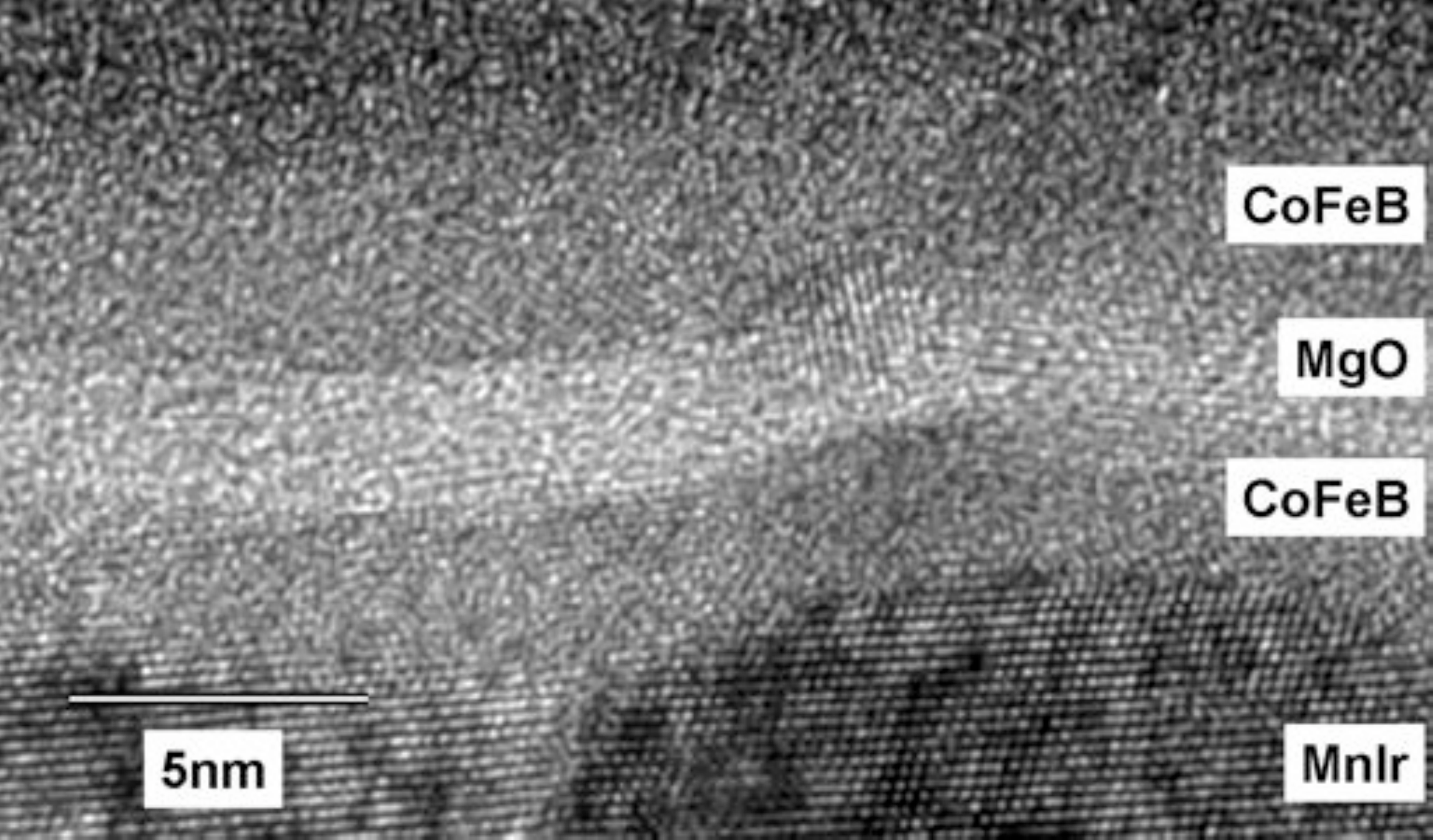}
\caption{\label{fig:intact_zoom} HRTEM micrograph of the crystalline MgO barrier sandwiched between two amorphous CoFeB layers obtained from a working TMR element.  }
\end{figure}
The high-resolution electron micrograph of the barrier region obtained from the same MTJ clearly reveals the MgO barrier sandwiched between two CoFeB electrodes (Figure \ref{fig:intact_zoom}). The latter show typical contrasts of amorphous materials. It should be noted that also on this scale the barrier is homogeneous without any indication of pinholes. 

\begin{figure}
\includegraphics[width=80mm]{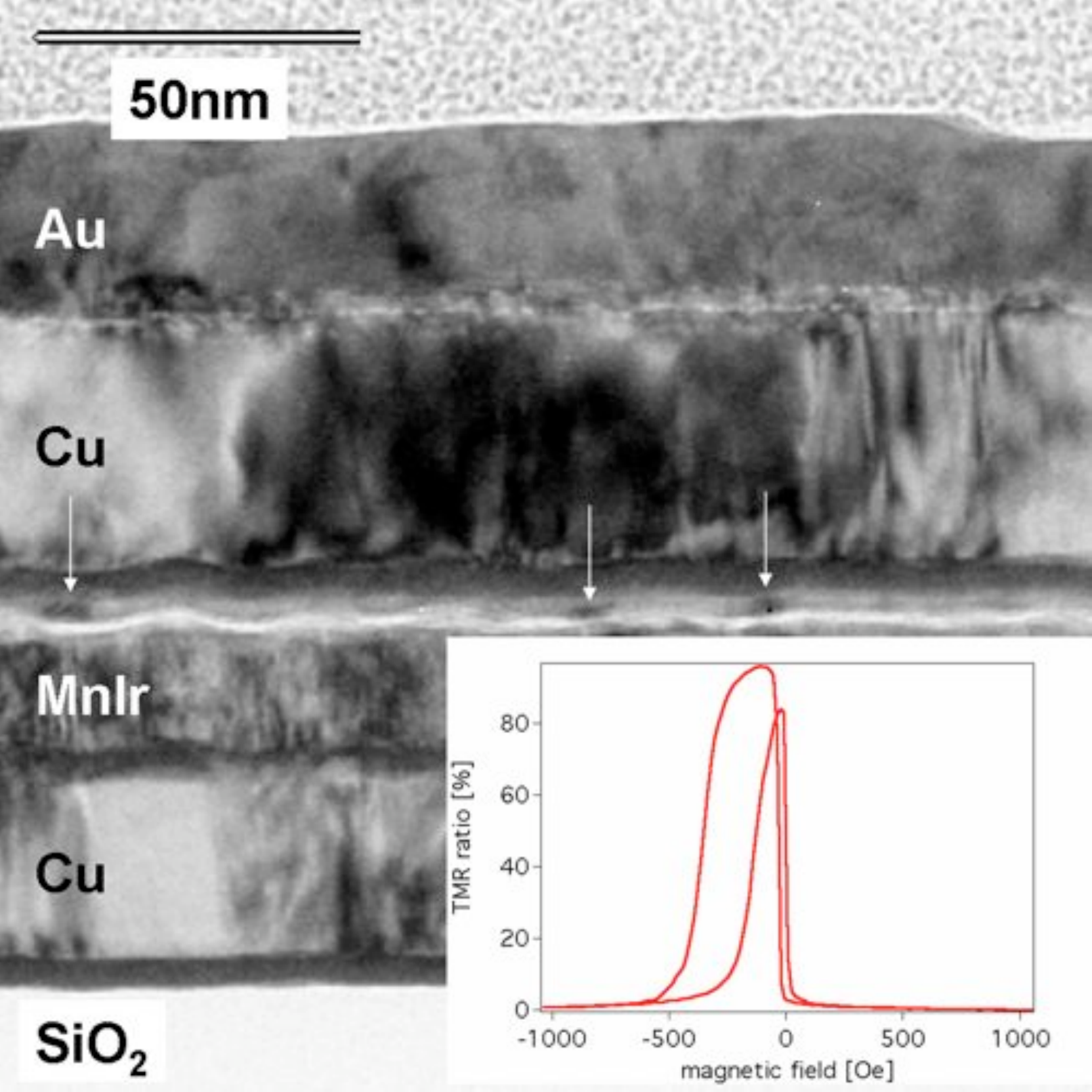}
\caption{\label{fig:broken_overview}Low-magnification TEM micrograph in cross-section geometry obtained from a TMR element after dielectric breakdown due to high voltage stresses. On this scale, local regions of the MgO tunneling barrier appear to be thinner compared to their neighbourhood; in addition, crystalline diffraction contrast in the CoFeB can be seen at these spots (white arrows). The inset shows the corresponding $R$ vs.\ $H$ loop of the same junction before focussed ion beam thinning and before voltage stress.}
\end{figure}
That picture changes if a voltage stress is applied to the junctions leading to a dielectric breakdown of the junction. The inset in Figure \ref{fig:broken_overview} displays the magnetoresistance loops of another junction before a voltage of 1.5\,V stressed the junction through a dielectric breakdown. The junction had very similar behavior before the breakdown and a substantially suppressed TMR-ratio afterwards. 
Compared to the initial situation, structural changes are clearly visible in the TEM already on a 10nm scale (Fig. \ref{fig:broken_overview}). White arrows indicate regions where quite strong diffraction contrast is locally observed 
in the upper CoFeB electrode which provides evidence that the material is at least partially crystallized. Furthermore, the MgO barrier appears to be broken at such spots. 

\begin{figure}
\includegraphics[width=80mm]{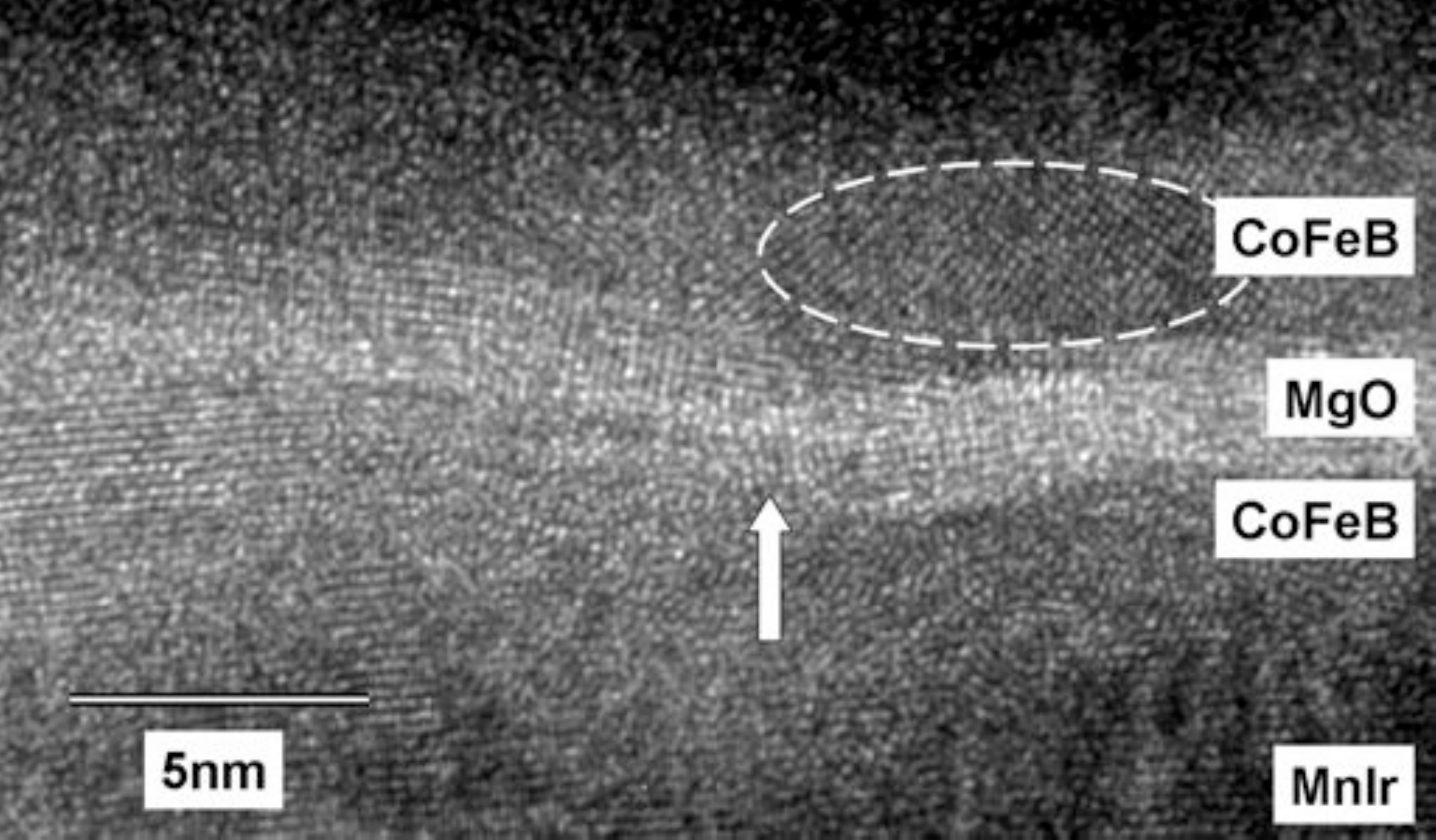}
\caption{\label{fig:broken_zoom}Transmission electron microscopy image of the tunnel barrier region of a voltage stressed magnetic tunnel junction. The white arrow indicates a pinhole in the MgO layer. In addition, a region of crystalline CoFeB is shown in the region enclosed by the dahed line.}
\end{figure}

One of these breaks is shown in Figure \ref{fig:broken_zoom} which is a HRTEM micrograph from such a damaged region. In the center of the image a locally interrupted MgO layer is revealed in addition to a crystalline grain in the top CoFeB electrode (dahed region). 

The breakdown in magnetic tunnel junctions is often explained by means of the E-model, which was originally proposed by McPherson and Mogul for $\rm SiO_2$ \cite{jap1998V84S1513}. In that model, the electric field and therefore the voltage is the important parameter that leads to the initial breakdown of the tunnel barrier (the first pinhole). This holds true here, since the breakdown voltage for the alumina as well as the MgO junctions are both in the order of 1.5\,V. On the other hand, the current flowing through the MgO is 200 times higher than in alumina devices. The (area-) resistances of the prepared MgO magnetic tunnel junctions is $\rm 54\,k\Omega \mu m^2$, the earlier produced alumina MTJs showed $\rm \approx\,10\,M\Omega \mu m^2$. 

This large current difference might explain the occurrence of many pinholes in the presented MgO case  after the first pinhole has formed (and single pinholes in the case of alumina). On order to do this we use the model proposed by H.\ Xi et.\ al.\ \cite{mmm2007V319S60} based on Kolmogorov \cite {iansm1937V3S355} and Avrami \cite{jcp1940V8S212}. They assume that constant stress of the junction by an electrical current leads to a constant pinhole formation rate $n_0$. If pinholes do not grow in size after they appear, the area ratio $a_p$ being covered with pinholes in respect to time $t$ is then given by
\begin{equation}
a_p(t) = 1- \exp \left( { - c n_0 t} \right)
\end{equation}
where $c$ is a constant. If we assume that the pinhole formation rate $n_0$ is higher if the stress current $j$ is higher, then we directly get $a_p^{\rm MgO}(t)  > a_p^{\rm AlO_x}(t)$, since the argument in the exponential function is larger. Assuming $n_0 \propto j$ the MgO pinholes appear 200 times faster than $\rm AlO_x$ pinholes.

We can also further illustrate that behavior, if we assume that the formation of a pinhole reduces the current flux (stress) in a surrounding area due to the electron flux through the conductive pinhole: The size of the affected area depends on the resistance of the tunnel barrier. The higher the tunnel resistance the larger the affected area will be. Therefore we can expect a stronger decrease of electrical stress in the case of the appearance of a single pinhole for higher tunnel resistances as it was the case for $\rm AlO_x$ based junctions.

%
In summary, we prepared magnetic tunnel junctions with MgO tunnel barrier to investigate the dielectric breakdown by transmission electron microscopy. All junctions were characterized by transport measurements. Then, one half of the junctions was stressed by voltages of about 1.5\,V to induce a dielectric breakdown. After that, both junction types were prepared by focussed ion beam to get transmission electron microscopy samples out of the actual junctions. We found that the breakdown in MgO junctions leads to a large number of breaks in the tunnel barrier and compared this to earlier investigations of Alumina based junctions.
\begin{acknowledgments}
We gratefully acknowledge J. Schmalhorst for helpful discussions. 
\end{acknowledgments}


\end{document}